\documentclass[pra,twocolumn,floats,showpacs,preprintnumbers,tighten,epsfig,superscriptaddress]{revtex4}

\usepackage{graphicx}
\usepackage{dcolumn}
\usepackage{bm}
\usepackage{amsmath}
\usepackage{amsfonts}
\usepackage{amssymb}

\begin{document}

\title{Interacting classical and quantum particles}

\author{Alvin J.~K.~Chua}
\author{M.~J.~W.~Hall}
\author{C.~M.~Savage}
\affiliation{
Research School of Physics and Engineering, Australian National University, Canberra ACT
0200, Australia}

\date{\today {}}

\begin{abstract}
We apply Hall and Reginatto's theory of interacting classical and quantum ensembles to harmonically coupled particles, with a view to understanding its experimental implications.  This hybrid theory has no free parameters and makes distinctive predictions that should allow it to be experimentally distinguished from quantum mechanics. It also bears on the questions of quantum measurement and quantum gravity.
\end{abstract}

\maketitle
\preprint{Version: Draft \today }

Does physics contain a classical sector, such as gravity? How does the quantum sector react back on the measurement devices that collapse it? What is the correct theoretical description of nano-mechanical systems in which quantum and classical objects are coupled? Such questions motivate the investigation of theories of interacting quantum and classical systems.

A number of candidate theories have been developed, but most have been found wanting in some aspect \cite{Salcedo,Peres and Terno,Boucher}. In this paper we investigate the ensemble theory of Hall and Reginatto \cite{Hall and Reginatto PRA, Hall PRA, Reginatto and Hall JP}, which avoids the difficulties in certain other attempts to unite quantum and classical dynamics within a unified mathematical structure.

Experimental investigation of objects straddling the boundary between the quantum and the classical is becoming possible with nano-mechanical techniques \cite{Painter arxiv,Bowen,Romero-Isart}, such as coupling of cantilevers or membranes to Bose-Einstein condensates \cite{Treutlein}, and with electromagnetic systems, such as SQUID rings coupled to classical circuits \cite{Everitt}. In so far as this boundary is experimentally unexplored, a quantum mechanical description of such experiments \cite{Steinke} should be compared against the foil of an alternative theory. 

Amongst the best developed alternatives to quantum mechanics are the collapse theories that postulate a transition between quantum and classical mechanics at a scale determined by a free parameter \cite{Adler, Bassi}. In contrast, the hybrid theory of Hall and Reginatto postulates distinct quantum and classical sectors, not necessarily a transition between them. It is a different kind of theory to collapse theories, and in particular it has no free parameters. Its predictions may be compared with those of quantum mechanics and of collapse theories.

There are difficulties with a standard quantum field theory of gravity \cite{Penrose book}, and a number of lines of research are seeking a consistent quantisation method. Although there are good reasons to want to quantise gravity \cite{Kiefer book}, classical gravity has not yet been ruled out \cite{Diosi,Callender,Carlip,Feynman book}, despite one particular theory  having been experimentally disproved \cite{Page experiment}. One of the obstacles to embracing classical gravity is the lack of well-developed theories encompassing both the quantum and classical sectors. The hybrid theory of Hall and Reginatto achieves this and has been applied to the case of a classical gravitational wave interacting with a quantized scalar field \cite{Albers}, providing a counterexample to claims that this is not possible \cite{DeWitt, Eppley}.

Some formulations of quantum mechanics require a classical sector to accommodate measurement devices. The lack of a quantitative theoretical distinction between the classical and quantum sectors is one aspect of the quantum measurement problem \cite{Schlosshauer RMP}. Indeed, axiomatisations of quantum mechanics usually include measurements as an unanalyzed notion \cite{Chiribella}. Another aspect of the measurement problem is that within standard theory, classical systems act on quantum systems, by reduction of the state, but quantum systems do not react back on classical systems. This back action is normally negligible when the measurement apparatus is large compared to the quantum system. However, at the quantum-classical boundary this may no longer be true.

Besides their fundamental interest, theories of interacting classical and quantum systems may be useful for developing consistent approximations of subsystems as classical, for example in quantum chemistry \cite{Prezhdo}.

\begin{figure}[b]
\includegraphics[width=8cm]{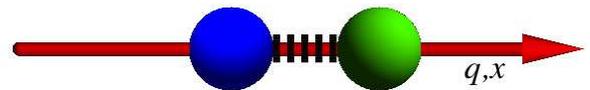}
\caption{(color online) Schematic illustration of the physical system considered in this paper. The spheres represent the particles, and the arrow the single dimension to which they are confined. The dashed line between the particles represents their harmonic interaction. Otherwise, they are free to move.}
\label{fig1}
\end{figure}
In the following pages, we apply the theory of Hall and Reginatto, which we shall refer to as the hybrid theory, to a classical particle interacting via a harmonic potential with a quantum particle, Fig.~\ref{fig1}. Our purpose is to understand the predictions of the theory and how they compare to the familiar cases of two interacting classical particles or two interacting quantum particles. In particular, we investigate the coupling between the relative and centre of mass degrees of freedom that is characteristic of the hybrid theory \cite{Hall and Reginatto PRA,Reginatto and Hall JP}, and find distinctive experimental signatures of hybrid dynamics.

We now outline the hybrid theory. A full account may be found in the papers of Hall and Reginatto \cite{Hall and Reginatto PRA,Hall PRA,Reginatto and Hall JP}. In this paper, we restrict our attention to the simplest case of one spatial dimension, with one classical particle and one quantum particle. Let the quantum coordinate be $q$ and the classical coordinate $x$. The theory is based on probability densities in configuration space. Let the probability density for finding the quantum particle at $q$ and the classical particle at $x$ be $P(q,x)$. Specifying the dynamics of $P(q,x)$ in the Hamiltonian formalism requires another function $S(q,x)$ that is canonically conjugate to $P(q,x)$, and an ensemble Hamiltonian
\begin{eqnarray} \label{vqx}
H[P,S] = \int dq\,dx\, P\,\left[ \frac{(\partial_q S)^2}{2m_q}
+ \frac{(\partial_x S)^2}{2m_x}  \right. \nonumber\\
\left. + V(q,x,t) + \frac{\hbar^2}{4} \frac{(\partial_q \ln P)^2}{2m_q} \right] .
\end{eqnarray}
The masses of the quantum and classical particles are $m_q$ and $m_x$ respectively.  The interaction potential  between the particles is $V(q,x,t)$. We define the Poisson bracket \cite{Hall PRA,goldstein} between two functionals $A[P,S]$ and $B[P,S]$ as
\begin{equation} \label{poiss} 
\{A,B\} = \int dq\,dx\, \left( \frac{\delta A}{\delta P} \frac{\delta B}{\delta S} -\frac{\delta B}{\delta P} \frac{\delta A}{\delta S}  \right)  .
\end{equation}
For a functional $L[f] = \int dz F(f,\partial_z f, z)$, for example, the variational derivative has the usual form \cite{goldstein}
\begin{equation}
\delta L/ \delta f = \partial F/\partial f - \partial_z [\partial F/\partial (\partial_z f)] .
\end{equation}
Using this, the dynamical equation for any functional $A[P,S,t]$ is \cite{Hall PRA}
\begin{equation} \label{evo} 
d A/d t = \{A,H\} +\partial A/\partial t .
\end{equation} 
In particular, the dynamics of $P$ and $S$ are given by
\begin{eqnarray} \label{P}
\frac{\partial P}{\partial t} &=&  - \frac{1}{m_q} \partial_q \left( P  \partial_qS  \right) 
- \frac{1}{m_x} \partial_x \left(P \partial_xS \right) , \\
 \label{S}
\frac{\partial S}{\partial t} &=& - \frac{1}{2m_q}  (\partial_qS)^2 -  \frac{1}{2m_x}  (\partial_xS)^2- V \nonumber \\ 
&&+ \frac{\hbar^2}{2m_q}\frac{\partial_q^2 P^{1/2}}{P^{1/2}} .
\end{eqnarray}
The final term may be recognized as the quantum potential \cite{Bohm}. These equations may be combined by defining the complex function $\psi(q,x) =\sqrt{P(q,x)} \; e^{iS(q,x)/\hbar}$. Eqs. (\ref{P},\ref{S}) are then equivalent to
\begin{eqnarray} \label{se}
i\hbar \frac{\partial \psi}{\partial t} &=& - \frac{\hbar^2}{2m_q}\partial_q^2\psi  + V \psi \nonumber \\
&& - \frac{\hbar^2}{2m_x}\partial_x^2\psi 
 + \frac{\hbar^2}{2m_x} \frac{\psi}{| \psi |} \partial_x^2| \psi |  .
\end{eqnarray}
The first three terms on the right hand side of this equation have the familiar Schr\"{o}dinger equation form. The final,  nonlinear, term ensures that the particle with coordinate $x$ is classical \cite{Holland Book, Rosen 64, Rosen 85, Schiller}. 

Observables in the hybrid theory are defined to be as similar as possible to those in quantum and classical physics. They are ensemble expectation values, defined by straightforward extensions of the usual definitions \cite{Hall and Reginatto PRA, Hall PRA, Hall JPA}. For any real classical phase space function $f(x,p_x)$,  the corresponding expectation value is
\begin{equation} \label{cf}  
\langle f \rangle = \int dq\,dx\, P \, f(x,\partial_x S)  .
\end{equation}
For any Hermitian operator $M$ associated with the quantum particle, the corresponding  expectation value is
\begin{equation} \label{qm}
\langle M \rangle = \int dq\,dx\, \psi^* M\psi . 
 \end{equation}
Expectation values of products of classical and quantum observables may be defined based on various principles; for example, the exact uncertainty principle \cite{Hall GRG}. We define them so that various straightforward principles give consistent results. A simple prescription is to first promote classical quantities to quantum ones, use the quantum definition Eq.~(\ref{qm}), and then take the classical limit of the promoted quantity by setting $\hbar \rightarrow 0$ only where it originates from the classical observable \cite{Hall and Reginatto PRA}. The only such products we will use are
\begin{equation} \label{productsx}
\langle qx \rangle = \int dq\,dx\, P \, q \, x , 
 \end{equation}
and
\begin{equation} \label{productsp}
\langle p_q  p_x \rangle = \int dq\,dx\, P \, \partial_q S \, \partial_x S ,
 \end{equation}
which have the same form as if both particles were classical. The second also follows from the symmetrized observable $\frac{1}{2} \langle p_x p_q + p_q p_x \rangle$ by the straightforward procedure of using the classical form $\partial_x S$ for $p_x$ and the quantum operator $-i \hbar \partial_q$ for $p_q$.

The hybrid theory has previously been applied to find analytic expressions for coherent states of the hybrid harmonic oscillator with the interaction potential \cite{Hall and Reginatto PRA} 
\begin{equation} \label{spring} 
V(q,x,t) = \frac{1}{2} k ( q-x )^2 .  
\end{equation}
In this paper, we extend that analysis to the more general squeezed Gaussian states. Hence their covariances become time-dependent rather than static. By making ansatzes, we reduce the partial differential equations Eqs.(\ref{P},\ref{S}) to ordinary differential equations that are straightforward to solve numerically.

We formulate the problem in matrix-vector form with vector coordinates $\xi =(q,x)^T$ \cite{Hall and Reginatto PRA}. The dynamical equations (\ref{P}) and (\ref{S}) are then
\begin{eqnarray} \label{posc}
\frac{\partial P}{\partial t} &+& \nabla \cdot \left(PU\nabla S\right) = 0 , \\
\frac{\partial S}{\partial t} &+& \frac{1}{2}(\nabla S)^T U\nabla S + \frac{1}{2}\xi^T C\xi \nonumber \\
 &-& \frac{\hbar^2}{2}\frac{\nabla \cdot (E_i U E_i \nabla  \sqrt{P})}{\sqrt{P}} =0 , \label{sosc}
\end{eqnarray}
where $\nabla = (\partial_q , \partial_x )^T$, and we have used the matrices
\begin{equation} \label{block}
C = k  \left(
 \begin{array}{cc} 1& -1 \\ -1 & 1 \end{array} 
\right),~
U = \left(
\begin{array}{cc} \frac{1}{m_q} & 0\\0 & \frac{1}{m_x} \end{array} 
\right) .
\end{equation}
Depending on the choice of the matrix $E_i$, this form of the equations applies to either the classical $E_c$, quantum $E_q$, or hybrid $E_h$ case:
\begin{equation} \label{E}
E_h = \left(
\begin{array}{cc} 1 & 0\\0 & 0 \end{array} \right),~
E_q = \left(
\begin{array}{cc} 1 & 0\\0 & 1 \end{array} \right),~
E_c = \left(
\begin{array}{cc} 0 & 0\\0 & 0 \end{array} \right) .
\end{equation}
We make the Gaussian ansatz for $P$:
\begin{equation} \label{pgauss} 
P(\xi,t) = \frac{\sqrt{|K|}}{2\pi} e^{-\frac{1}{2} (\xi -\alpha)^T K(\xi -\alpha)}, 
\end{equation}
where $K$ (the inverse of the covariance matrix) is a positive-definite symmetric matrix, $|K|$ is its determinant, and $\alpha$ is a vector.  This is consistent with the equations of motion if and only if $S$ is at most quadratic in the coordinates $\xi$:
\begin{equation} \label{slin} 
S(\xi,t) = \frac{1}{2} (\xi -\alpha)^T L  (\xi -\alpha)+ \beta^T (\xi-\alpha) + \sigma  , 
\end{equation}
where $L$ is a symmetric matrix, $\beta$ is a vector and $\sigma$ is a scalar. Substituting these ansatzes into Eqs.~(\ref{posc}) and (\ref{sosc}) yields a system of ordinary differential equations \cite{Hall and Reginatto PRA}. The terms quadratic in $\xi -\alpha$ yield \cite{useful results}:
\begin{eqnarray} \label{k}
\frac{dK}{dt} + KUL +LUK = 0,
\\
\label{l}
\frac{dL}{dt} + LUL +C - \frac{\hbar^2}{4} K E_i U E_i K = 0,
\end{eqnarray}
where the equations have been written to emphasise the symmetry of the matrices. The terms linear in $\xi -\alpha$ yield:
\begin{equation} \label{ab}
\frac{d \alpha}{dt} -U\beta = 0, ~~~\frac{d \beta}{dt} + C\alpha = 0 .
\end{equation}
These equations are the same as for a classical oscillator, and have the solution
\begin{eqnarray} \label{asol}
\alpha &=& (a +b t) \left(
 \begin{array}{c} 1 \\ 1 \end{array} 
 \right) + \nonumber \\
&& c \mu \cos \left( \omega_\mu t+\phi\right)
 \left(
 \begin{array}{c} m_q^{-1} \\ -m_x^{-1} \end{array} 
\right) ,
\\ \label{bsol}
\beta &=& b 
 \left(
 \begin{array}{c} m_q \\ m_x \end{array} 
 \right) 
  -c \mu \, \omega_\mu \sin \left( \omega_\mu t +\phi\right)
 \left(
 \begin{array}{c} 1 \\ -1 \end{array} 
 \right)  ,
\end{eqnarray}
where $a$, $b$,  $c$ and $\phi$ are determined by the initial conditions. $\mu = m_q m_x /(m_q + m_x )$ is the reduced mass, and $\omega_{\mu}=\sqrt{k/\mu}$ is the oscillator frequency. Hence the probability distribution $P$ is centred on the classical motion, as in the quantum case. The zeroth-order terms in $\xi -\alpha$ yield a differential equation for $\sigma$, which plays no further role in our analysis.

The differences between the probability distribution dynamics of the classical, the quantum, and the hybrid cases are  found in the covariance matrix $Z=K^{-1}$. The elements of $Z$ are related to the coordinate covariances by:
\begin{eqnarray}
Z_{qq} &=& {\rm Var}(q),~~Z_{xx}={\rm Var}(x) , \nonumber \\
Z_{qx} &=& {\rm Cov}(q,x) =
\langle qx\rangle - \langle q\rangle \langle x\rangle .
\end{eqnarray}

We have numerically solved the nonlinear differential equations (\ref{k},\ref{l}) for the matrices $K$ and $L$ \cite{Mathematica}. For our initial conditions, we assume that the two particles are uncorrelated, such that  $K$ and the covariance matrix $Z$ are initially diagonal.  The momentum covariances may be expressed as
\begin{eqnarray}
{\rm Cov}(p_j,p_k) &=& \langle p_j p_k \rangle - \langle p_j\rangle \langle
p_k \rangle \nonumber \\
&=& L Z L + \frac{\hbar^2}{4}E_i K E_i .
\end{eqnarray}
In the quantum case, we make $S$ initially linear in the coordinates so that the matrix $L=0$. The quantum particles are then in a minimum-uncertainty state.  In the hybrid case, we give the classical particle the same momentum variance as the quantum particle by taking: $L_{xx}(0)=0.5 \hbar K_{xx}, \; L_{qq}(0)= L_{xq}(0)=0$. We do not present solutions for the classical case as the solutions for $K$ and $L$ are singular at a quarter-period, although there are no singularities in the observables. Our present approach is not suited to the classical case, although it works well for the hybrid and quantum cases.
\begin{figure}[tbp]
\includegraphics[width=8cm]{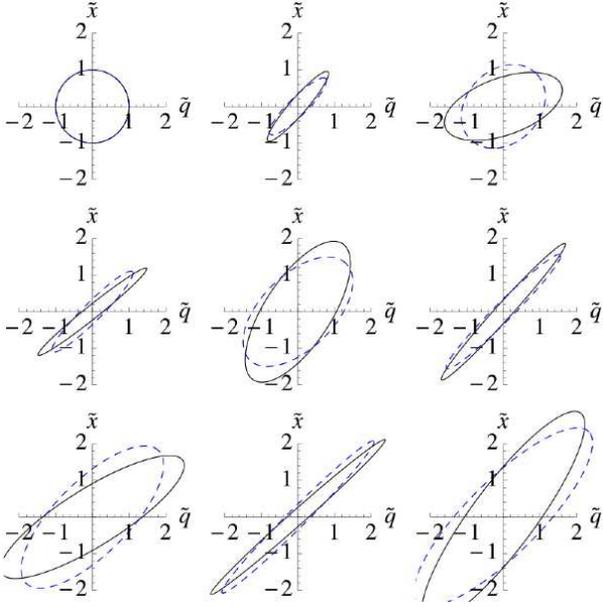}
\caption{(color online) Error ellipses for the covariance matrix $Z$, shown at quarter-period intervals over two oscillation periods. The solid black ellipses are for the hybrid case and the dashed blue ellipses are for the quantum case.  The axes are the dimensionless coordinates: $\tilde{q}=q/x_0$ and $\tilde{x}=x/x_0$, where $x_0$ is the harmonic oscillator length $x_0 = (\hbar/ \mu \omega_\mu)^{1/2}$.  First row: $t=0,T/4,T/2$. Second row: $t=3T/4,T,5T/4$. Third row: $t=3T/2,7T/4,2T$. Here $T = 2 \pi /\omega_\mu$ is the oscillator period. The initial conditions are given in the text, with $K_{xx} = K_{qq} = 1/ x_0^2$. In units of the reduced mass $\mu$, the masses are: $m_q = m_x =  2 \mu$.}
\label{fig2}
\end{figure}

Fig.~\ref{fig2} shows the time dependence of the error ellipses associated with the position covariance matrix $Z$, for both hybrid and quantum cases, for equal masses $m_q = m_x$. These ellipses have their axes in the directions of the unit eigenvectors of $Z$, with the semi-axis lengths equal to the square roots of the corresponding eigenvalues. Hence the ellipses' orientations may be interpreted as giving the directions of maximum and minimum coordinate variance, with the semi-axis lengths giving the values of the square roots of those variances.

The covariance ellipses in Fig.~\ref{fig2} have similar orientations,  with the
quantum ellipses strictly aligned with the center-of-mass and relative
coordinate directions:
\begin{equation} \label{cr coords}
R = \frac{m_q q +m_x x}{M},~~~ r = q-x ,
\end{equation}
where $M = m_q + m_x$ is the total mass. Prominent shared features are:
variance oscillations in the relative direction, and variance growth in the
center-of-mass direction.  However, a significant difference between the
two cases is the rocking of the orientation of the hybrid ellipses, in
contrast to the fixed alignment of the quantum ellipses.  This rocking
arises from a coupling between the center-of-mass and relative coordinates,
which underpins the signatures of hybrid dynamics considered in this
paper.

\begin{figure}[tbp]
\includegraphics[width=8cm]{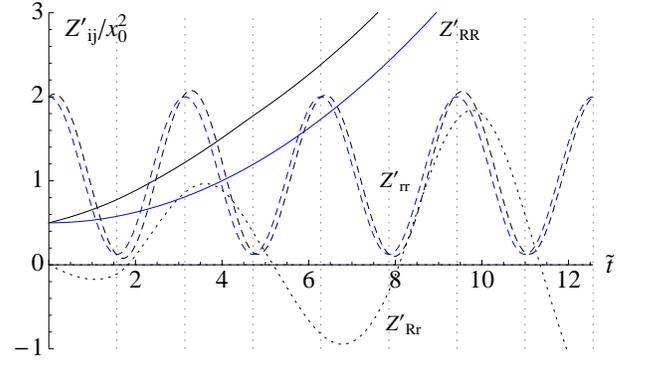}
\caption{(color online) Coordinate covariances in center-of-mass and relative coordinates versus dimensionless time $\tilde{t} = t \omega_\mu$ for hybrid (black curves) and quantum (blue curves) oscillators. The solution is the same as in  Fig.~\ref{fig2}. Solid lines: $ {\rm Var}(R) = Z'_{RR}$. Dashed lines:  ${\rm Var}(r) = Z'_{rr}$. Dotted line:  ${\rm Cov}(R,r) = Z'_{Rr}$. In the quantum case, ${\rm Cov}(R,r) = 0$.}
\label{fig3}
\end{figure}

We next consider the solution in center-of-mass $R$ and relative $r$ coordinates. We will find that the center-of-mass has distinctive dynamics in the hybrid case.  The associated momenta are the usual:
\begin{equation} \label{cr momenta}
p_R = p_q + p_x,~~~ p_r = \frac{m_x}{M}p_q - \frac{m_q}{M}p_x .
\end{equation}
Defining the new coordinate vector $\xi' =(R, r)^T$, the coordinate transformation matrix $T$ such that $\xi = T \xi'$ is
\begin{equation} \label{T matrix}
T =  \left( \begin{array}{cc} 1& m_x / M \\ 1 & -m_q / M
\end{array}  \right) .
\end{equation}
In the center-of-mass and relative coordinates, the correlation matrix $Z'$ is given by
\begin{equation} \label{Z cr }
Z' = (T^T K T)^{-1} .
\end{equation}
Fig.~\ref{fig3} shows the components of $Z'$ for the equal mass case of Fig.~\ref{fig2}. It shows that the spreading is occurring in the centre-of-mass coordinate (solid lines), while the relative coordinate variance is oscillating (dashed lines). For the quantum case, the centre-of-mass variance has the quadratic dependence on time characteristic of a free particle \cite{Tannor},  confirming our methodology. The most notable difference between the hybrid and quantum cases is the behaviour of the correlation between the relative and center-of-mass coordinates $Z'_{Rr}$. In the quantum case, the coordinates decouple and there is no correlation: $Z'_{Rr}(t) = 0$; in the hybrid case they are coupled, and Fig.~\ref{fig3} shows a growing oscillation.

Such coupling has been previously noted, as the ensemble Hamiltonian Eq.~(\ref{vqx}) in center-of-mass and relative coordinates is \cite{Hall and Reginatto PRA}
\begin{eqnarray} \label{Hcr}
&& H[P,S] = \int dR \,dr \, P\,\left[ \frac{(\partial_{R} S)^2}{2M} \right.
 \nonumber \\
&+& \left. \frac{\hbar^2 m_q}{8M^2} (\partial_{R} \ln P)^2 \right]
 \nonumber \\
&+&  \int dR \,dr \, P\,\left[ \frac{(\partial_{r} S)^2}{2 \mu} +  \frac{\hbar^2}{8m_q} (\partial_{r} \ln P)^2 +V(r) \right]
\nonumber \\
&-&  \frac{\hbar^2}{4M} \int dR \,dr \, \frac{ (\partial_{R} P)( \partial_{r} P) }{P} .
\end{eqnarray}
In particular, the last term couples the center-of-mass and relative coordinates. 

The center-of-mass kinetic energy $E_R$, relative kinetic energy $E_r$ and the interaction energy $\langle V  \rangle$ are given by:
\begin{eqnarray} \label{Energies }
E_R = \frac{\langle p_R^2 \rangle}{2 M} = \frac{1}{2M} ( \langle p_q^2 \rangle + \langle p_x^2 \rangle +2  \langle p_q p_x \rangle )
, \\
E_r =  \frac{\langle p_r^2 \rangle}{2 \mu} = \frac{1}{2M} ( \frac{m_x}{m_q} \langle p_q^2 \rangle + \frac{m_q}{m_x} \langle p_x^2 \rangle -2  \langle p_q p_x \rangle )
, \\
 \langle V  \rangle = \frac{k}{2} \langle r^2 \rangle = \frac{k}{2} \langle (q-x)^2 \rangle .
\end{eqnarray}
The total energy $E = E_R  + E_r + \langle V \rangle$ and the center-of-mass momentum $\langle p_R \rangle$ are conserved, as for the classical and quantum cases.  The classical oscillator parts of the solution, Eqs.~(\ref{asol},\ref{bsol}), contribute $\frac{1}{2} M b^2$ to the center-of-mass kinetic energy, $\frac{k}{2} c^2 \sin^2(\omega_\mu t + \phi)$ to the relative kinetic energy, and $\frac{k}{2} c^2 \cos^2(\omega_\mu t + \phi)$ to the interaction energy. The sum of all these classical energies is independently conserved. That our numerical solutions of Eqs.~(\ref{k},\ref{l}) conserve all the required observables in both quantum and hybrid cases is a check on their validity. 

\begin{figure}[tbp]
\includegraphics[width=8cm]{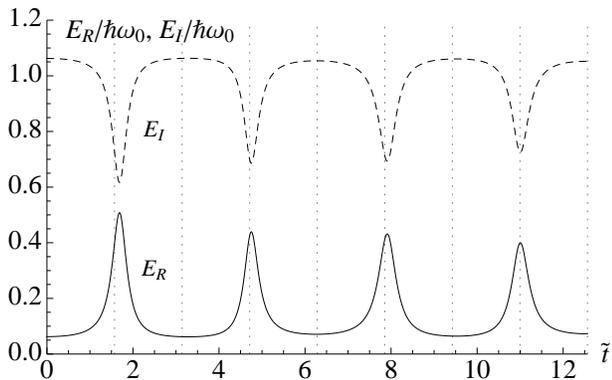}
\caption{Hybrid oscillator center-of-mass kinetic energy $E_R$ (solid line) and internal energy $E_I = E_r + \langle V \rangle$ (dashed line) versus dimensionless time $\tilde{t} = t \omega_\mu$. Masses are equal, $m_q = m_x$. The total energy is $E = 1.125 \hbar \omega_\mu$.The vertical dotted lines are at quarter-periods. The contribution of the classical motion to the internal energy has not been included. The solution is the same as in  Fig.~\ref{fig2}. In the case of quantum or classical physics $E_R$ and $E_I$ are constants.}
\label{fig4}
\end{figure}

In both the classical and quantum cases the internal energy, $E_I = E_r + \langle V \rangle$, and the center-of-mass kinetic energy, $E_R$, are also separately conserved. In the hybrid case this is no longer true, as can be seen in Fig.~(\ref{fig4}). In this case, approximately 40\% of the total energy is exchanged between the internal and center-of-mass degrees of freedom. The peaks of the center-of-mass kinetic energy occur near quarter-periods of the harmonic motion. This surprising behaviour is due to the final coupling term in the ensemble Hamiltonian Eq.~(\ref{Hcr}). Alternatively, it may be attributed to the final nonlinear term in the modified Schr\"{o}dinger equation (\ref{se}). The nonlinearity explains why the dynamics are non-periodic.

The distinctive hybrid dynamics, in which the center-of-mass kinetic energy is time-dependent, are the central result of this paper, and might be used to experimentally test the theory. To do so we must first find classical particles to experiment on. The predictions of the hybrid theory give clues as to how such particles might be identified.  

There are no free parameters in the hybrid theory. Once the experimental parameters, such as masses, are known, the hybrid dynamics are quantitatively and absolutely determined. If those dynamics are not observed, the theory is refuted. 

Experimental signatures of hybrid dynamics can be found in position or momentum measurements. The most dramatic examples are perhaps in the dynamics of the center-of-mass kinetic energy $E_R$, shown in Fig.~\ref{fig4} for equal masses. Ensembles of momentum measurements allow $E_R$ to be inferred, either directly if the center-of-mass momentum is measured, or indirectly if the individual particle momenta are measured.  Momenta might be measured using scattering experiments. Both quantum and classical systems have constant center-of-mass kinetic energy, while only in hybrid systems does it vary in time. Hence the observation of a time-varying center-of-mass kinetic energy would be evidence for hybrid dynamics. 

Hybrid systems could also be distinguished from quantum or classical systems by ensembles of position measurements. For example, Fig.~\ref{fig3} shows that in the equal mass case, only in the hybrid dynamics does an initially zero covariance between the center-of-mass and relative coordinates evolve to differ from zero.

Our numerical solutions may be related to specific experimental parameters through the reduced mass $\mu$ and oscillator frequency $\omega_\mu$. The length scale $x_0 = (\hbar/ \mu \omega_\mu)^{1/2}$ and energy scale $\hbar \omega_\mu$ follow.

\begin{figure}[tbp]
\includegraphics[width=8cm]{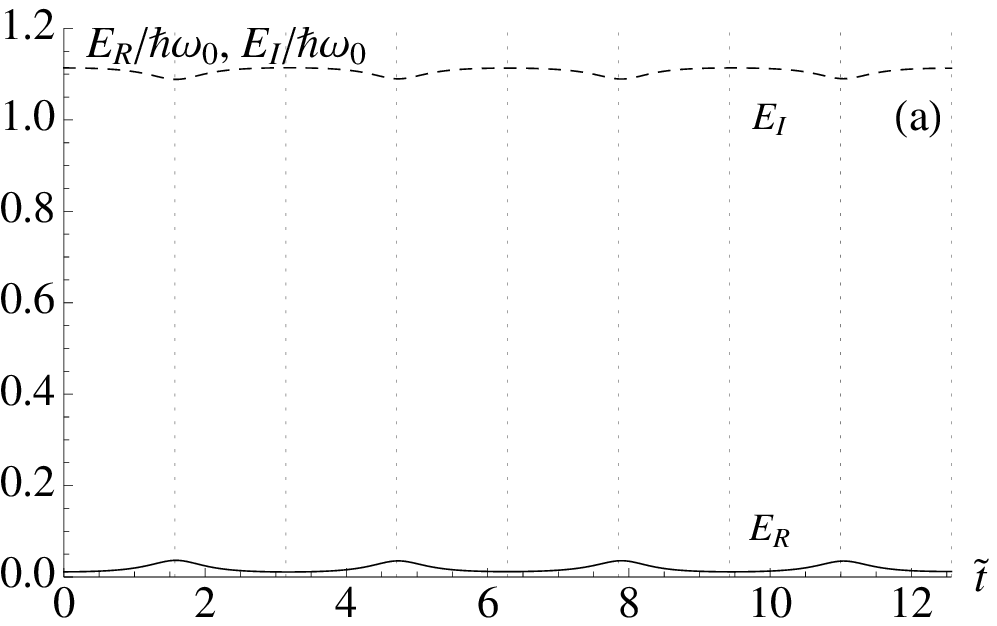}
\includegraphics[width=8cm]{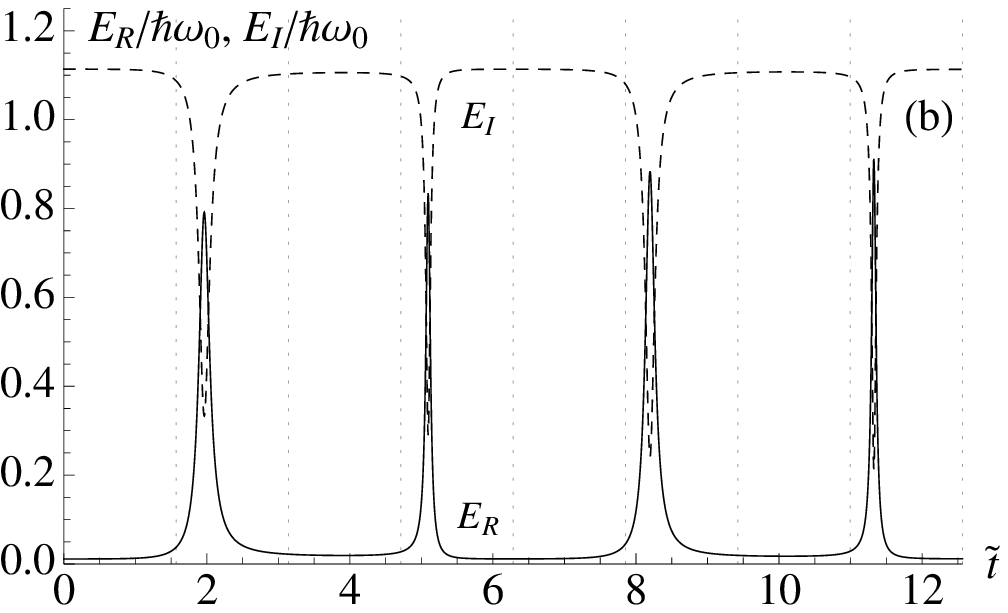}
\caption{Hybrid oscillator center-of-mass kinetic energy $E_R$ (solid line) and internal energy $E_I = E_r + \langle V \rangle$ (dashed line) versus dimensionless time $\tilde{t} = t \omega_\mu$. (a) Dominant classical mass $m_x = 20 m_q = 21 \mu$. (b) Dominant quantum mass $m_q = 20 m_x = 21 \mu$. The vertical dotted lines are at quarter-periods. The contribution of the classical motion to the internal energy has not been included. Initial conditions and other parameters are the same as for Figs.~\ref{fig2}, \ref{fig3} and \ref{fig4}.}
\label{fig5}
\end{figure}

We next consider cases in which the classical and quantum particles have different masses. Fig.~\ref{fig5} shows the center-of-mass and internal energies in two cases, with either the classical, Fig.~\ref{fig5}(a), or quantum, Fig.~\ref{fig5}(b), mass dominant. Due to decoherence it is likely to prove easier to keep a lighter object quantum-mechanical; hence the case with dominant classical mass, Fig.~\ref{fig5}(a), is more likely to be the first to be realised experimentally. In that case, the exchange of energy between the center-of-mass kinetic energy $E_R$ and the internal energy $E_I$ is less than in the equal-mass case. The dynamics then better approximate those of conventional quantum or classical physics, and the hybrid case is less distinctive. 

In the case with quantum mass dominant, Fig.~\ref{fig5}(b), the exchange of energy exceeds that in the equal-mass case, and the signature of hybrid dynamics is clearer. As the ratio of the classical mass to the quantum mass becomes smaller, the peaks in the center-of-mass kinetic energy seen in Fig.~\ref{fig5}(b) become narrower.

\begin{figure}[tbp]
\includegraphics[width=8cm]{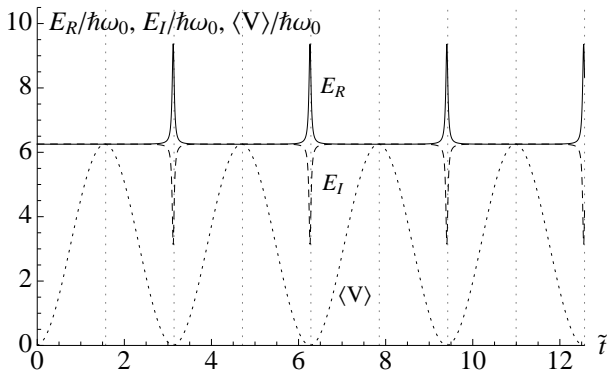}
\caption{Hybrid oscillator center-of-mass kinetic energy $E_R$ (solid line), internal energy $E_I = E_r +\langle V \rangle$ (dashed line), and potential energy $\langle V \rangle$ (dotted line) versus dimensionless time $\tilde{t} = t \omega_\mu$. Masses are equal. The total energy is $E = 12.51 \hbar \omega_\mu$.  The vertical dotted lines are at quarter-periods. The contribution of the classical motion to the internal energy has not been included. Initial conditions and other parameters are the same as for Figs.~\ref{fig2}, \ref{fig3} and \ref{fig4}, except $K_{xx} = K_{qq} = 100 / x_0^2$.}
\label{fig6}
\end{figure}

In the specific cases examined so far, the total energy has been slightly greater than $\hbar \omega_\mu$. Fig.~\ref{fig6} shows the center-of-mass kinetic, internal and potential energies when the total energy is about $12 \hbar \omega_\mu$. This was achieved by reducing the initial coordinate variances, and hence increasing the initial momentum variances, by factors of $100$. When the potential energy is near zero, the center-of-mass kinetic energy spikes while the internal energy dips to conserve the total energy. Similar spiking behaviour is also seen in Figs.~\ref{fig4} and \ref{fig5}. The spiking also occurs in the kinetic energy of the individual quantum particle, at the expense of a dip in the classical particle's kinetic energy. However, the total kinetic energy oscillates smoothly with no spiking. This behaviour is specific to the hybrid physics, and cannot be explained in terms of classical or quantum dynamics.

In summary, we have applied the hybrid theory to the illustrative case of two harmonically coupled particles. In this case, we were able to use ansatzes to reduce the theory's partial differential equations to ordinary differential equations that are easy to solve numerically. The hybrid theory predicts different dynamics to classical physics or quantum physics. These differences are due to the distinctive coupling of the relative and center-of-mass degrees of freedom, and to the nonlinearity of the hybrid theory. Further work is required to develop an understanding of the new hybrid dynamics. 

Applications of the theory to more realistic cases, such as  nano-mechanical systems, will require numerical solutions of the theory's dynamical partial differential equations, Eqs.~(\ref{P},\ref{S}). An attractive feature of the hybrid theory is its lack of free parameters, which allows experiments to absolutely confirm or refute it.


\begin{thebibliography}{99}
\bibitem{Mathematica} Numerical solutions were performed with Mathematica 8 using the NDSolve function with default settings.
\end{thebibliography}
\end{document}